# KEA: Practical Automatic Keyphrase Extraction

*Ian H. Witten,[*] Gordon W. Paynter,[*] Eibe Frank,[*] Carl Gutwin[†] and Craig G. Nevill-Manning[‡]*

[*] Dept of Computer Science,
University of Waikato,
Hamilton, New Zealand.
{ihw,gwp,eibe}@cs.waikato.ac.nz

[†] Dept of Computer Science,
University of Saskatchewan,
Saskatoon, Canada
gutwin@cs.usask.ca

[‡] Dept of Computer Science,
Rutgers University,
Piscataway, New Jersey
nevill@cs.rutgers.edu

**ABSTRACT**

Keyphrases provide semantic metadata that summarize and characterize documents. This paper describes Kea, an algorithm for automatically extracting keyphrases from text. Kea identifies candidate keyphrases using lexical methods, calculates feature values for each candidate, and uses a machine-learning algorithm to predict which candidates are good keyphrases. The machine learning scheme first builds a prediction model using training documents with known keyphrases, and then uses the model to find keyphrases in new documents. We use a large test corpus to evaluate Kea's effectiveness in terms of how many author-assigned keyphrases are correctly identified. The system is simple, robust, and publicly available.

**INTRODUCTION**

Keyphrases provide a brief summary of a document's contents. As large document collections such as digital libraries become widespread, the value of such summary information increases. Keywords and keyphrases[1] are particularly useful because they can be interpreted individually and independently of each other. They can be used in information retrieval systems as descriptions of the documents returned by a query, as the basis for search indexes, as a way of browsing a collection, and as a document clustering technique.

In addition, keyphrases can help users get a feel for the content of a collection, provide sensible entry points into it, show how queries can be extended, facilitate document skimming by visually emphasizing important phrases; and offer a powerful means of measuring document similarity (e.g. [6], [8], [13]).

Keyphrases are usually chosen manually. In many academic contexts, authors assign keyphrases to documents they have written. Professional indexers often choose phrases from a predefined "controlled vocabulary" relevant to the domain at hand. However, the great majority of documents come without keyphrases, and assigning them manually is a tedious process that requires knowledge of the subject matter. Automatic extraction techniques are potentially of great benefit.

Several methods have been proposed for generating or extracting summary information from text (e.g. [1], [7], [10]). In the specific domain of keyphrases, there are two fundamentally different approaches: *keyphrase assignment* and *keyphrase extraction*. Both use machine learning methods, and require for training purposes a set of documents with keyphrases already attached.

Keyphrase assignment seeks to select the phrases from a controlled vocabulary that best describe a document. The training data associates a set of documents with each phrase in the vocabulary, and builds a classifier for each phrase. A new document is processed by each classifier, and assigned the keyphrase of any model that classifies it positively (e.g. [3]). The only keyphrases that can be assigned are ones that have already been seen in the training data.

Keyphrase extraction, the approach used here, does not use a controlled vocabulary, but instead chooses keyphrases from the text itself. It employs lexical and information retrieval techniques to extract phrases from the document text that are likely to characterize it [12]. In this approach, the training data is used to tune the parameters of the extraction algorithm.

---

[1] throughout this document we use the latter term to subsume the former

| Protocols for secure, atomic transaction execution in electronic commerce | | Neural multigrid for gauge theories and other disordered systems | | Proof nets, garbage, and computations | |
|---|---|---|---|---|---|
| anonymity | *atomicity* | disordered systems | disordered | *cut-elimination* | cut |
| *atomicity* | *auction* | tems | gauge | linear logic | *cut elimination* |
| *auction* | customer | *gauge fields* | *gauge fields* | *proof nets* | garbage |
| *electronic commerce* | *electronic commerce* | *multigrid* | interpolation kernels | sharing graphs | *proof net* |
| privacy | intruder | neural multigrid | nels | typed lambda-calculus | weakening |
| real-time | merchant | neural networks | length scale | calculus | |
| *security* | protocol | | *multigrid* | | |
| *transaction* | *security* | | smooth | | |
| | third party | | | | |
| | *transaction* | | | | |

This paper describes a new keyphrase extraction algorithm, Kea, that is simple and effective, and performs at the current state of the art [5]. It uses the Naïve Bayes machine learning algorithm for training and keyphrase extraction. An implementation is available from the New Zealand Digital Library project (http://www.nzdl.org/).

Kea's output is illustrated in Table 1, which shows the titles of three research articles and two sets of keyphrases for each article. One set gives the keyphrases assigned by the author; the other was determined automatically from the article's full text. Phrases in common between the two sets are italicized.

In each case, the author's keyphrases and the automatically-extracted keyphrases are quite similar, but it is not too difficult to guess which phrases are the author's. The giveaway is that Kea, in addition to choosing several good keyphrases, also chooses some that authors are unlikely to use—for example, *gauge*, *smooth*, and especially *garbage*! Despite these anomalies, the automatically-extracted lists seem to provide a reasonable description of the three papers. In the case where no author-specified keyphrases were available, Kea's choices would be a valuable resource to someone encountering these three articles for the first time.

Our goal, therefore, is to provide useful metadata where none existed before. Although we evaluate Kea's performance by comparing with the author's own keyphrases, we do not expect to equal them. If we can extract reasonable summaries from text documents, we give a valuable tool to the designers and users of digital libraries. The remainder of this paper describes Kea. The next section details the design of the algorithm. We then give an example of the prediction model generated by Kea and show how it is used to assess a candidate keyphrase. Following that, we report on several experiments designed to test Kea's effectiveness and to explore the effects of varying parameters in the extraction process.

## THE KEA ALGORITHM

Kea's extraction algorithm has two stages:

1. Training: create a model for identifying keyphrases, using training documents where the author's keyphrases are known.

2. Extraction: choose keyphrases from a new document, using the above model.

The process is outlined in Figure 1. Both stages choose a set of *candidate phrases* from their input documents, and then calculate the values of certain attributes (called features) for each candidate. We describe these two steps first, and then outline the training and extraction stages in more detail.

### Candidate phrases

Kea chooses candidate phrases in three steps. It first cleans the input text, then identifies candidates, and finally stems and case-folds the phrases.

#### Input cleaning

ASCII input files are filtered to regularize the text and determine initial phrase boundaries. The input stream is split into tokens (sequences of letters, digits and internal periods), and then several modifications are made:

- punctuation marks, brackets, and numbers are replaced by phrase boundaries;
- apostrophes are removed;
- hyphenated words are split in two;
- remaining non-token characters are deleted, as are any tokens that do not contain letters.

The result is a set of lines, each a sequence of tokens containing at least one letter. Acronyms containing periods, like *C4.5*, are retained as single tokens.

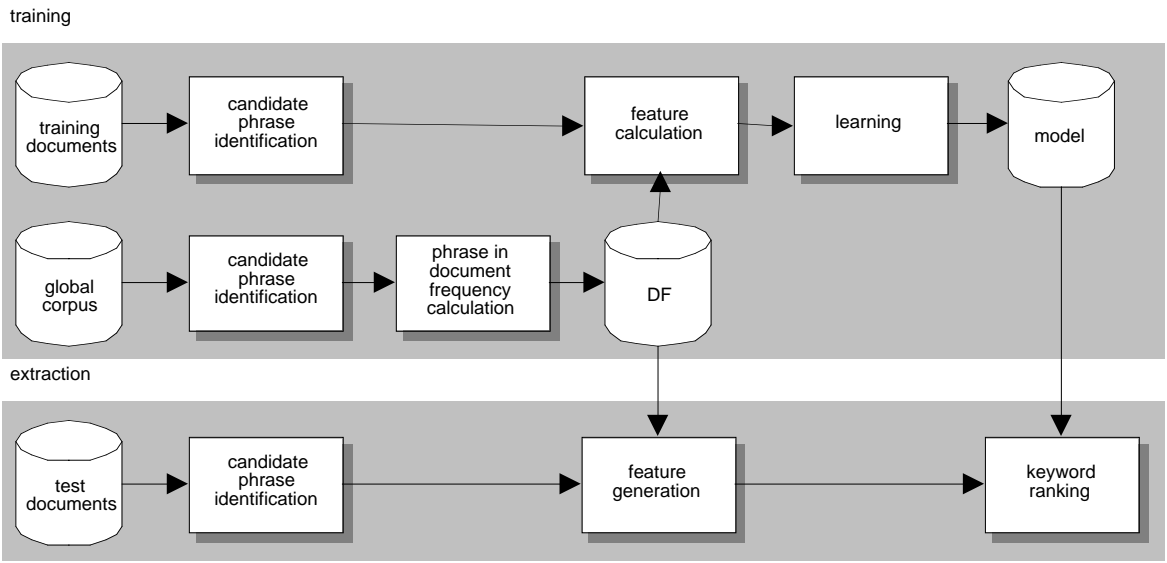

**Figure 1**  The training and extraction processes

*Phrase identification*
Kea then considers all the subsequences in each line and determines which of these are suitable candidate phrases. We have investigated several methods for determining suitability, such as looking for noun phrases, but we have found that the following rules are both simple and effective:
1. Candidate phrases are limited to a certain maximum length (usually three words).
2. Candidate phrases cannot be proper names (i.e. single words that only ever appear with an initial capital).
3. Candidate phrases cannot begin or end with a stopword.

The stopword list contains 425 words in nine syntactic classes (conjunctions, articles, particles, prepositions, pronouns, anomalous verbs, adjectives, and adverbs). For most of these classes, all the words listed in an on-line dictionary were added to the list. However, for adjectives and adverbs, we introduced several subclasses, and words from the subclasses were added only if they overlapped the sixty most common words in the Brown corpus [9]. Furthermore, we only added frequently-occurring words from these subclasses.

All contiguous sequences of words in each input line are tested using the three rules above, yielding a set of candidate phrases. Note that subphrases are often candidates themselves. Thus, for example, a line that reads *the programming by demonstration method* will generate *programming*, *demonstration*, *method*, *programming by demonstration*, *demonstration method*, and *programming by demonstration method* as candidate phrases, because *the* and *by* are on the stopword list.

*Case-folding and stemming*
The final step in determining candidate phrases is to case-fold all words and stem them using the iterated Lovins method. This involves using the classic Lovins stemmer [11] to discard any suffix, and repeating the process on the stem that remains until there is no further change. So, for example, the phrase *cut elimination* becomes *cut elim*.

Stemming and case-folding allow us to treat different variations on a phrase as the same thing. For example, *proof net* and *proof nets* are essentially the same, but without stemming they would have to be treated as different phrases. In addition, we use the stemmed versions to compare Kea's output to the author's keyphrases. We consider an author-specified keyphrase to have been successfully identified if, when stemmed, it is the same as a machine-generated keyphrase, also stemmed. That is why in Table 1 the phrases *cut-elimination* and *cut elimination*, and *proof nets* and *proof net*, are considered equivalent.

We retain the unstemmed words for each phrase, in their original capitalization, for presentation to the user in case the phrase does turn out to be a keyphrase. When several different capitalizations occur, the most frequent version is chosen.

**Feature calculation**
Two features are calculated for each candidate phrase and used in training and extraction. They are: *TF×IDF*, a measure of a phrase's frequency in a document compared to its rarity in general use; and *first occurrence*, which is the distance into the document of the phrase's first appearance.

*TF×IDF*
This feature compares the frequency of a phrase's use in a particular document with the frequency of that phrase in general use. General usage is represented by *document frequency*—the number of documents containing the phrase in some large corpus. A phrase's document frequency indicates how common it is (and rarer phrases are more likely to be keyphrases). Kea builds a document frequency file for this purpose using a corpus of about 100 documents. Stemmed candidate phrases are generated from all documents in this corpus using the method described above. The document frequency file stores each phrase and a count of the number of documents in which it appears.

With this file in hand, the TF×IDF for phrase *P* in document *D* is:

$$\text{TF} \times \text{IDF} = \frac{\text{freq}(P,D)}{\text{size}(D)} \times -\log_2 \frac{\text{df}(P)}{N}, \text{ where}$$

1. freq(*P*,*D*) is the number of times *P* occurs in *D*
2. size(*D*) is the number of words in *D*
3. df(*P*) is the number of documents containing *P* in the global corpus
4. *N* is the size of the global corpus.

The second term in the equation is the log of the probability that this phrase appears in any document of the corpus (negated because the probability is less than one). If the document is not part of the global corpus, df(*P*) and *N* are both incremented by one before the term is evaluated, to simulate its appearance in the corpus.

*First occurrence*
The second feature, first occurrence, is calculated as the number of words that precede the phrase's first appearance, divided by the number of words in the document. The result is a number between 0 and 1 that represents how much of the document precedes the phrase's first appearance.

*Discretization*
Both features are real numbers and must be converted to nominal data for the machine-learning scheme. During the training process, a discretization table for each feature is derived from the training data. This table gives a set of numeric ranges for each feature, and values are replaced by the range into which the value falls. Discretization is accomplished using the supervised discretization method described in [4].

**Training: building the model**
The training stage uses a set of training documents for which the author's keyphrases are known. For each training document, candidate phrases are identified and their feature values are calculated as described above. To reduce the size of the training set, we discard any phrase that occurs only once in the document. Each phrase is then marked as a keyphrase or a non-keyphrase, using the actual keyphrases for that document. This binary feature is the *class feature* used by the machine learning scheme.

The scheme then generates a model that predicts the class using the values of the other two features. We have experimented with a number of different machine learning schemes; Kea uses the Naïve Bayes technique (e.g [2]) because it is simple and yields good results. This scheme learns two sets of numeric weights from the discretized feature values, one set applying to positive ("is a keyphrase") examples and the other to negative ("is not a keyphrase") instances. An example model is described in Section 3.

**Extraction of new keyphrases**
To select keyphrases from a new document, Kea determines candidate phrases and feature values, and then applies the model built during training. The model determines the overall probability that each candidate is a keyphrase, and then a post-processing operation selects the best set of keyphrases.

When the Naïve Bayes model is used on a candidate phrase with feature values *t* (for TF×IDF) and *d* (for distance), two quantities are computed:

$$P[yes] = \frac{Y}{Y+N} P_{TF \times IDF}[t \mid yes] \, P_{distance}[d \mid yes] \qquad (1)$$

and a similar expression for P[*no*], where *Y* is the number of positive instances in the training files—that is, author-identified keyphrases—and *N* is the number of negative instances—that is, candidate phrases that are not keyphrases. (The Laplace estimator is used to avoid zero probabilities. This simply replaces *Y* and *N* by *Y*+1 and *N*+1.)

The overall probability that the candidate phrase is a keyphrase can then be calculated:

$$p = P[yes] / (P[yes]+P[no]) \qquad (2)$$

Candidate phrases are ranked according to this value, and two post-process steps are carried. First, TF×IDF (in its pre-discretized form) is used as a tie-breaker if two phrases have equal probability (common because of the discretization). Second, we remove from the list any phrase that is a subphrase of a higher-ranking phrase. From the remaining ranked list, the first *r* phrases are returned, where *r* is the number of keyphrases requested.

## KEYPHRASE EXTRACTION EXAMPLE

To illustrate the Naïve Bayes modeling method, we exhibit a model for keyphrase extraction that was learned in one experiment, and show its application to a particular phrase.

### Sample model

Table 2 shows the model. For this training set, TF×IDF was discretized into five fixed levels, and first occurrence into four levels. The discretization boundaries are given at the top of Table 2.

Using this discretization, there are nine feature weights for positive examples and nine for negative ones. For example, $P_{TF \times IDF}[1 \mid yes]$ is the proportion of positive examples that have a discretized TF×IDF value of 1. The values learned for these weights are shown in the middle of Table 2.

The final component of the learned model is the number of positive and negative instances in the training set, shown at the bottom of Table 2. These determine the prior probability of a candidate phrase being a keyphrase, in the absence of any other information.

### Application of the model

As an example of keyphrase assignment, the phrase *cut elimination*, with stem *cut elim*, appears 16 times in the third paper of Table 1. The size of this paper is 5114 words; the phrase first appears at word 130. There are 132 documents in the global corpus, and *cut elim* appears in just one, but this paper is not in the global corpus, so these counts are incremented by 1. This gives *cut elim* the feature values TF×IDF = 0.0189, distance = 0.0254. After discretization, these become 4 and 3.

The a posteriori likelihoods of this phrase being in the *yes* and *no* classes are calculated from Equation (1), and the overall probability for it being a keyphrase is calculated from Equation (2) as 0.0805. This makes it the fifth candidate phrase in the probability ordered list, so it will be returned as a keyphrase provided five or more are requested.

The individual words *cut* and *elim* are also candidate phrases. Although *cut* has the same probability as *cut elimination*, it is ranked higher because its (undiscretized) TF×IDF is greater; thus it will also appear as a keyphrase. On the other hand, *elim* will never be chosen as a keyphrase, no matter how many are sought, because its probability is lower than that of its superphrase.

## EVALUATION

We carried out an empirical evaluation of Kea using documents from the New Zealand Digital Library. Our goals were to assess Kea's overall effectiveness, and also to investigate the effects of varying several parameters in the extraction process. We measured keyphrase quality by counting the number of matches between Kea's output and the keyphrases that were originally chosen by the document's author. The following sections outline our experimental methodology and report the results.

### Methodology

*Procedure*

Kea was evaluated using the Computer Science Technical Reports (CSTR) collection of the NZDL. From the 46,000 documents in this corpus, we chose 1800 where the author had supplied keyphrases. From these 1800, we randomly chose a test set of 500 documents, leaving 1300 as a pool from which to select training documents. The large test set reduces measurement error, so our results will closely approximate the expected values for any particular document. Finally, a further set of documents were chosen at random from the remainder of the CSTR as our global corpus, used to build the document-frequency file.

| *Discretization table* | Feature | Discretization ranges | | | | |
|---|---|---|---|---|---|---|
| | | 1 | 2 | 3 | 4 | 5 |
| | TF×IDF | < 0.0031 | [0.0031, 0.0045) | [0.0045, 0.013) | [0.013, 0.033) | ≥ 0.033 |
| | distance | < 0.0014 | [0.0014, 0.017) | [0.017, 0.081) | ≥ 0.081 | |
| *Class probabilities* | Feature | Values | Discretization ranges | | | | |
| | | | 1 | 2 | 3 | 4 | 5 |
| | TF×IDF | P[TF×IDF \| *yes*] | 0.2826 | 0.1002 | 0.2986 | 0.1984 | 0.1182 |
| | | P[TF×IDF \| *no*] | 0.8609 | 0.0548 | 0.0667 | 0.0140 | 0.0036 |
| | distance | P[distance \| *yes*] | 0.1952 | 0.3360 | 0.2515 | 0.2173 | |
| | | P[distance \| *no*] | 0.0194 | 0.0759 | 0.1789 | 0.7333 | |
| *Prior probabilities* | Class | Training instances | | Prior probability | | |
| | *yes* | 493 | | P(*yes*) = Y/(Y+N) = 0.0044 | | |
| | *no* | 112183 | | P(*no*) = N/(Y+N) = 0.9956 | | |

We then carried out four experiments to determine:
- Kea's overall effectiveness
- the effect of changing the size and source of the global corpus
- the effect of changing the number of training documents
- Kea's performance using abstracts rather than full text

Results from each of these experiments are given below; first, however, we describe our quality measures, and discuss the advantages and disadvantages of using author-specified keyphrases as a standard.

*Measures*
We assess Kea's effectiveness by counting the keyphrases that were also chosen by the document's author, when a fixed number of keyphrases are extracted. We use this measure instead of the more common information-retrieval metrics of *precision* and *recall* for three reasons. First, a single overall value is more easily interpreted than two values. Second, precision and recall can be misleading, for it is easy to maximize precision at the expense of recall (by returning the single most promising candidate phrase), or recall at the expense of precision (by returning all candidates). Third, our measure fits well with the expected behaviour of end-users, who will likely ask for a certain number of keyphrases for a document. If required, however, precision can be calculated by dividing our measure by the number of phrases retrieved.

We chose to measure Kea against the choices of the document's author for several reasons: this method of evaluation is simple, can be carried out automatically, and allows the comparison of different extraction schemes. However, there are several disadvantages to using author keyphrases as a standard—primarily that authors do not always choose keyphrases that best describe the content of their paper. Authors might choose phrases to slant their work a certain way, or to maximize its chance of being noticed by particular searchers. Also, keyphrases are often chosen hastily, just before a document is finalized. Finally, one can argue that authors are in any case poorly qualified to choose phrases to describe their work for others.

This problem raises two issues. First, the variance in author choices makes it more difficult for an automatic extraction scheme to perform well. Second, Kea's incorrect choices (those that did not match an author choice) are not necessarily poor keyphrases. A more revealing approach might be to use human judges to independently assess the quality of Kea's phrases, without using the original author's choices at all. This approach, however, requires considerable resources even for a single experiment, and so we leave this method for future studies.

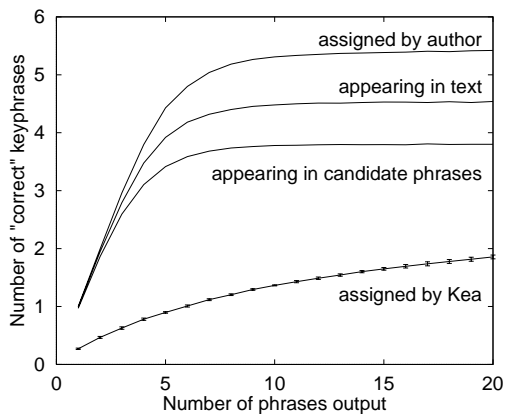

**Figure 2** Overall performance

### Results
*Overall effectiveness*
Our first experiment assessed Kea's overall effectiveness, when extracting up to 20 keyphrases per test document. This experiment used 50 training documents, the standard 500-document test set, and a global corpus of 100 documents. Selected results are shown in Table 3 below, and illustrated in Figure 2.

| Keyphrases extracted | Average matches with author keyphrases |
|---|---|
| 5 | 0.93 |
| 10 | 1.39 |
| 15 | 1.68 |
| 20 | 1.88 |

**Table 3** Overall performance

In Figure 2, the lowest line shows the average number of correct identifications. The upper lines show three limits on possible performance. The first shows how many keyphrases the author assigned: clearly it is not possible for any algorithm to do better than this using our measure of success. The asymptote shows that the test set has an average of 5.4 author-assigned keyphrases per document. The second line from the top indicates the number of keyphrases that appear in the document's text. No method of keyphrase *extraction* (as opposed to *assignment*) can possibly identify keyphrases that do not appear in the text. The third gives the number of keyphrases appearing within the *candidate* phrases (see Section 2.1).

Figure 2 thus illustrates where Kea loses ground. The difference between the two middle lines represents how many keyphrases are not selected by the candidate selection process. The difference between the bottom two lines represents how much better the machine learning scheme could conceivably do in finding the authors' keyphrases from among the candidates.

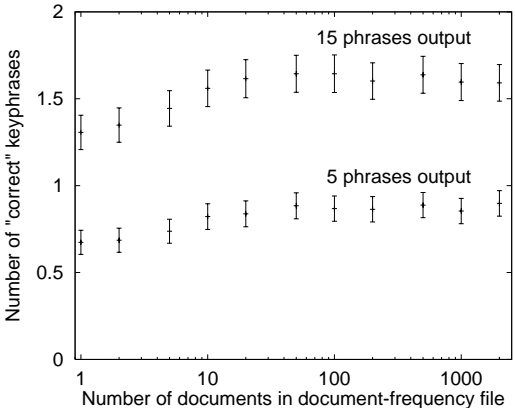

**Figure 3**  Effect of number of documents used when calculating TF×IDF

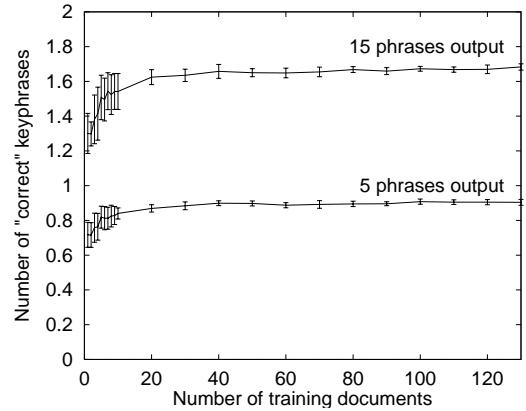

**Figure 4**  Performance against number of training files

The error bars on the lowest line (which are so small as to be barely visible) represent variance due to the choice of training documents. If one considers the population of all training sets of size 50, there is a 99% chance that the population mean lies within the error bar. Using training sets of only 50 documents represents the realistic situation where there are not many documents available with known keyphrases. Although the results for any given training set will differ, we can be 99% sure that Figure 2 accurately portrays the expected result over different training sets.

*Effect of size and source of global corpus*
We carried out a series of tests to determine how the size and source of the global corpus affects performance. As described in Section 2.2, the global corpus is used to build a document frequency file used in TF×IDF calculations. We were interested in the corpus' size since a larger global corpus will more closely approximate a phrase's true frequency in general use. We were also interested in the source of the global corpus' documents—in particular, whether the similarity of these documents to the test documents would affect performance.

To test the effect of the source, we built different global corpuses from: an independent set of similar documents, the training set, the training and test sets, the test set alone, and a set of documents containing a different kind of material. In our trials, no one global corpus significantly outperformed the others.

To test the effect of global corpus size, we tested Kea using corpuses of different sizes. For these trials, we used a training set of 130 documents, and the standard 500-document test set. All global corpuses were formed randomly from the CSTR documents without author-assigned keyphrases. As shown in Table 4 and in Figure 3, there is little to be gained by increasing the size of the global corpus beyond about ten documents, and after 50 documents, there is no further improvement. However, the document-frequency file is crucial for good results: without one, performance drops off dramatically.

| Documents in corpus | Average # matches (5 extracted) | Average # matches (15 extracted) |
| --- | --- | --- |
| 0 | ? | ? |
| 1 | 0.674 | 1.307 |
| 5 | 0.738 | 1.445 |
| 10 | 0.822 | 1.560 |
| 50 | 0.884 | 1.644 |
| 100 | 0.868 | 1.644 |
| 1000 | 0.854 | 1.596 |

**Table 4**  Effect of varying global corpus size

Figure 3 plots the number of keyphrases matched against the size of the global corpus. The error bars give 95% confidence intervals for the number of correct keyphrases extracted from a test document, given the particular training set.

*Effect of training set size*
Our third experiment investigated whether the number of training documents (those with keyphrases identified) affect performance. We were interested in the practical problem of how many training documents are necessary for good results. In this experiment, we use a standard global corpus of 100 CSTR documents, and the standard test set. We varied the size of the training set from 1 to 130 documents, and tested Kea's performance with each set.

Our results (Table 5 and Figure 4) show that performance improves steadily up to a training set of about 20 documents, and smaller gains are made until the training set holds 50 documents. Figure 4 plots the number of correctly-identified keyphrases, when 5 and 15 phrases are extracted, against the number of documents used for training. The error bars show 99% confidence limits.

| Training documents | Average # matches (5 extracted) | Average # matches (15 extracted) |
|---|---|---|
| 0 | 0.684 | 1.266 |
| 1 | 0.717 | 1.301 |
| 5 | 0.819 | 1.508 |
| 10 | 0.840 | 1.542 |
| 20 | 0.869 | 1.625 |
| 50 | 0.898 | 1.650 |
| 100 | 0.908 | 1.673 |

**Table 5** Effect of varying training set size

These results indicate that good extraction performance can be had with a relatively small set of training documents. In a real-world situation where a collection without any keyphrases is to be processed, human experts need only read and assign keyphrases to about 25 documents in order to extract keyphrases from the rest of the collection.

*Effect of document length*
Our final experiment considered whether Kea's performance suffers when it only uses the abstracts of documents to extract keyphrases, and compares it to performance on the full text. This experiment used the standard training, testing, and global corpus sets, except that documents with no abstract were ignored (leaving 110 training documents and 429 testing documents).

Table 6 shows the number of correct keyphrases extracted using both the short and full documents. As expected, Kea extracts fewer keyphrases from abstracts than from the full document text.

| Document length | Average # matches (5 extracted) | Average # matches (15 extracted) |
|---|---|---|
| Full text | 0.909 | 1.712 |
| Abstracts | 0.655 | 1.028 |

**Table 6** Effect of varying document length

Figure 5 plots curves for the short document trial only. The four solid lines, from top to bottom, indicate: the number of keyphrases assigned by the author, the number appearing in the shortened document, the number that appear in the candidate list, and the number that are correctly identified by Kea. The dashed line is the number of correct keyphrases identified when using the full document text. The main reason for the reduced performance when using abstracts seems to be that—not surprisingly—far fewer of the author's keyphrases appear in the abstract than can be found in the entire document.

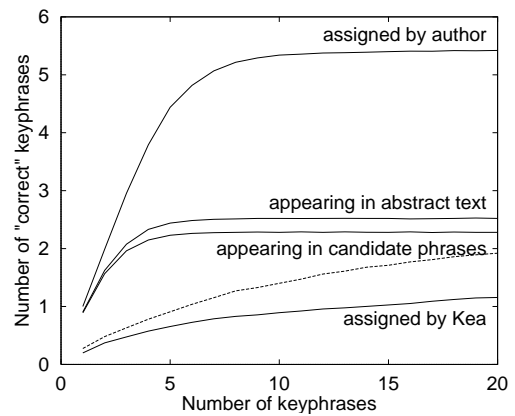

**Figure 5** Number of correct keyphrases against number of phrases extracted

## CONCLUSION

We have described and evaluated an algorithm for automatically extracting keyphrases from text. Our results show that Kea can on average match between one and two of the five keyphrases chosen by the author in this collection.[2] We consider this to be good performance. Although Kea find less than half the author's phrases, it must choose from many thousands of candidates; also, it is highly unlikely that even another human would select the same set of phrases as the original author.

Therefore, our next project is to leave the author's phrases behind and evaluate Kea's phrases with a more robust measure. We will use human judges to rate how well a set of extracted keyphrases summarize a particular document. Although this experiment will provide a more realistic assessment, it is clear that some of Kea's phrases are poor regardless of the measure. These poor phrases are not easy to weed out: the reason that *garbage* is a poor keyword (see Table 1) is subtle from a computational viewpoint. Therefore, we will also investigate techniques for determining what makes a phrase reasonable from a human perspective.

At present, Kea's performance is sufficient for the applications it was designed for: providing support for summarizing, browsing, searching and clustering in cases where manual keyphrase assignment is infeasible. It can and will greatly assist designers and users of large document collections.

Kea is available from the New Zealand Digital Library project (http://www.nzdl.org/).

---

[2] The version of Kea described here is domain-independent. Other experiments [5] show how performance can be improved by incorporating a degree of domain dependence.

**ACKNOWLEDGMENTS**

We would like to thank Peter Turney for sharing his datasets, discoveries, and experiences.